\documentclass[aps,prl,10pt, tightenlines, twoside, secnumarabic, 
superscriptaddress,showpacs, preprintnumbers, showpacs, nofootinbib, 
twocolumn]{revtex4-1}

\pdfoutput=1
\usepackage[english]{babel}
\usepackage{amsmath,amssymb,amsfonts, bm,bbm,slashed, subdepth}
\usepackage{graphicx}
\usepackage[sort&compress]{natbib}
\usepackage{xcolor}
\usepackage[normalem]{ulem}
\usepackage{hyperref}
\usepackage{cleveref}
\definecolor{red}{rgb}{1.0, 0, 0}
\usepackage{hyperref}
\usepackage{enumerate}
\usepackage{epsfig, subfigure}
\usepackage{setspace}
\usepackage{booktabs, tabularx}
\usepackage{units}

\hypersetup{
    pdfnewwindow=true,      
    colorlinks=true,       
    linkcolor=blue,          
    citecolor=blue,        
    filecolor=blue,      
    urlcolor=blue        
}

\allowdisplaybreaks

\bibpunct{[}{]}{,}{n}{}{,}

\newcommand{\chip}{\chi_{{}_+}}
\newcommand{\chim}{\chi_{{}_-}}
\newcommand{\chibp}{\bar{\chi}_{{}_+}}
\newcommand{\chibm}{\bar{\chi}_{{}_-}}

\newcommand{\sv}[0]{\langle \sigma v \rangle}

\begin{document}
\font\secret=cmr10 at 0.8pt
\title{Dark radiation alleviates problems with dark matter halos
}

\author{Xiaoyong Chu}
\email{xchu@ictp.it}
\affiliation{International Centre for Theoretical Physics, Strada Costiera 11, 34014 Trieste, Italy.}

\author{Basudeb Dasgupta}
\email{bdasgupta@ictp.it}
\affiliation{International Centre for Theoretical Physics, Strada Costiera 11, 34014 Trieste, Italy.}

\date{August 19, 2014}

\begin{abstract}
We show that a scalar and a fermion charged under a global $U(1)$ symmetry can not only explain the existence and abundance of 
dark matter (DM) and dark radiation (DR), but also imbue DM with improved scattering properties at galactic scales, while remaining consistent with all other observations. Delayed DM-DR kinetic decoupling eases the \emph{missing satellites} problem, while scalar-mediated self-interactions of DM ease the \emph{cusp vs. core} and \emph{too big to fail} problems. In this scenario, DM is expected to be pseudo-Dirac and have a mass $100\,{\rm keV}\lesssim m_\chi\lesssim 10\,{\rm GeV}$. The predicted DR may be measurable using the primordial elemental abundances from big bang nucleosynthesis (BBN), and using the cosmic microwave background (CMB).
\end{abstract}

\begin{flushright}
\end{flushright}

\maketitle

{\bf{\emph{Introduction.--}}} 
Cosmological and astrophysical data now firmly point towards the existence of new \emph{nonrelativistic} particles, dubbed dark matter (DM), and there is a vigorous experimental program underway to discover these particles and measure their properties. Dark radiation (DR), on the other hand, refers to new \emph{relativistic} particles that contribute to the cosmological energy density  but are otherwise decoupled from ordinary matter and radiation. There is neither clear evidence nor a definitive exclusion, but several independent analyses of cosmological data show tantalizing hints for DR\,\cite{Izotov:2010ca, Steigman:2014pfa, Giusarma:2014zza, Cheng:2014ota, Dvorkin:2014lea}, most recently to reconcile the results from the Planck Collaboration\,\cite{Ade:2013zuv} with those from BICEP2 Collaboration\,\cite{Ade:2014xna}.

Candidate particles for DR have been motivated by experimental results, e.g., additional neutrinos that explain oscillation anomalies, or to address conceptual problems in the visible sector, e.g., thermal axions that solve the strong CP problem. On the other hand, DR may have little to do with the observed particles in the Universe and instead may simply be light particles in the dark sector\,\cite{Blinnikov:1982eh, Foot:1991bp, Berezhiani:1995am, Pospelov:2007mp, Ackerman:mha}. Weinberg recently presented a set-up\,\cite{Weinberg:2013kea}, where the Goldstone bosons of a global symmetry in the dark sector lead to dark radiation, while the residual symmetry provides stability to a fermionic DM candidate. Its phenomenology has been explored in subsequent works\,\cite{Cheung:2013oya, Garcia-Cely:2013nin, Clarke:2013aya, Anchordoqui:2013bfa, Garcia-Cely:2013wda, Baek:2014awa}. In this \textit{Letter}, we show that if DM and DR share this common origin, it may naturally solve long-standing problems in DM structure-formation.

A weakly interacting massive particle explains the cosmological abundance of DM, but there are hints from observations of dwarf galaxies and the Milky-Way that something may be lacking in this description. N-body simulations of collisionless cold DM\,\cite{Navarro:1996gj} predict numerous dwarf satellite galaxies of the Milky-Way, that are not seen, viz., the \emph{missing satellites} problem\,\cite{Klypin:1999uc}. They also predict cuspy halos in dwarf galaxies where cores are observed\,\cite{Moore:1994yx, Flores:1994gz}, viz. the \emph{cusp vs. core} problem, and highly massive subhalos of Milky-Way-type galaxies that would be expected to host stars, but which aren't observed, viz., the \emph{too big to fail} problem\,\cite{BoylanKolchin:2011de}. It has been considered that inclusion of more detailed astrophysical processes\,\cite{MacLow:1998wv, Governato:2009bg, Silk:2010aw, VeraCiro:2012na, Wang:2012sv,Governato:2012fa, Brooks:2012ah, Zolotov:2012xd,Amorisco:2013uwa}, e.g., supernova feedback, low star-formation, tidal effects, etc., or new DM physics\,\cite{Spergel:1999mh, Bode:2000gq, Kaplinghat:2000vt, Boehm:2000gq, Sigurdson:2003vy, Hooper:2007tu, Kaplan:2009de}, e.g., self-interactions, warm DM, decays/annihilations, or DM-``baryon'' coupling etc., can solve \emph{some} of these problems. Exotic interactions between DM and ordinary matter, e.g., neutrinos\,\cite{Aarssen:2012fx} or sterile neutrinos\,\cite{Dasgupta:2013zpn} may be able to address \emph{all} these persistent problems. However, almost all models invoke additional physics specifically to address the small-scale problems.

We show that (i)~DM scattering off the DR bath, composed of the Goldstone bosons of the global symmetry, leads to delayed kinetic decoupling that erases the least massive DM halos, which can mitigate the \emph{missing satellites} problem, and (ii)~DM self-scattering mediated by the scalar mode leads to smoothening of the inner cusps of small galactic halos, which alleviates the \emph{cusp vs. core} and \emph{too big to fail} problems. Together, they can ease \emph{all} tensions between observations and cold DM simulations, with no need for any other particles or interactions. Simultaneously, the observed DM density and all other constraints can be satisfied, which predicts an observable abundance of~DR and the viable DM mass-range.

In the following, we demonstrate the above mechanism. We show how the small-scale problems may be addressed, and elaborate the consequences for DR and DM. We outline the phenomenology of this scenario, and conclude with a discussion and a summary of our results.

{\bf{\emph{Dark Matter and Dark Radiation.--}}} We consider the Lagrangian for the dark sector\,\cite{Weinberg:2013kea,Garcia-Cely:2013nin},
\begin{eqnarray}
{\mathcal L}_{\rm dark} &\ni &\partial_\mu \phi^* \partial^\mu \phi + \mu^2_\phi |\phi|^2 -\lambda_\phi |\phi|^4\notag\\
                 & +& i \bar \chi \gamma^\mu \partial_\mu \chi - M\bar \chi \chi - ({f_{\rm d}\over \sqrt{2} } \phi \chi^T C \chi +h.c.),\label{Wein:Lagrangian} 
\end{eqnarray}
where $\phi$ is a complex scalar and $\chi$ is a 4-component fermion, both charged under a global $U(1)$ symmetry. 
After symmetry-breaking, $\phi\equiv (v_\phi+\rho+i\eta)/\sqrt{2}\,$ has a vacuum expectation value $v_\phi$. Its CP-odd component $\eta$ becomes a massless Goldstone field while its CP-even component $\rho$ remains. At the same time, the last term in Eq.\,(\ref{Wein:Lagrangian}) splits the fermion field into two mass eigenstates $\chi_\pm$ with masses $m_{\chi_{_\pm}}= |M \pm f_{\rm d} v_\phi |$.
The obvious $Z_2$ residual symmetry, i.e., $\chi_{\pm}\rightarrow-\chi_{\pm}$ and $(\rho,\eta) \rightarrow (\rho,\eta)$, guarantees that the lighter mass eigenstate, which we take to be $\chim$, is stable, and therefore a viable DM candidate. Relativistic dark particles, e.g., the massless Goldstone mode $\eta$, yield DR.

We will be interested in DM and DR scattering processes mediated by the $\chi-\phi$ interaction in Eq.\,(\ref{Wein:Lagrangian}), which, after symmetry breaking, is rewritten as 
\begin{equation}
 -{f_{\rm d} \over 2}\rho(\chibm \chim-\,\chibp\chip) -{f_{\rm d}\over 2}\eta (\chibp\chim +\,\chibm\chip)\,.
\end{equation}
We will show that when $\chim$ and $\chip$ are quasi-degenerate, i.e., $m_{\chip}-m_{\chim} \equiv \Delta m_\chi \ll m_\chi$, the scattering processes can be appreciable and important. However, before we get to that, let's consider the cosmological abundances of DR and DM in this scenario.

The temperatures of the dark and the visible sectors are defined to be the temperatures of the bath of $\eta$ and photons (denoted by $\gamma$), respectively. We will assume that $T_{\star}$ is a temperature above which the dark sector was in thermal equilibrium with the visible sector. This may have been through processes common to both sectors at high-scale, e.g., inflaton decay. Below this temperature, the two sectors are decoupled but the conservation of entropy relates the temperatures in the two sectors as
\begin{equation}
 T_{\eta} = \left(\frac{g^*_{\rm d}(T_{\star})}{g^*_{\rm d}(T_{\eta})}\frac{g^*_{\rm v}(T_{\gamma})}{g^*_{\rm v}(T_{\star})}\right)^{1/3} T_{\gamma}\,\equiv \xi\,T_\gamma,
\end{equation}
where $g^*(T)$ are the effective number of relativistic degrees of freedom in the dark (${\rm d}$) and visible (${\rm v}$) sectors, respectively, at their temperatures $T$. In our model, $0\lesssim \xi(T_\gamma)\lesssim 1$. 

The DR density is given by relativistic particles in the dark sector, i.e., $\rho_{\rm DR}= \pi^2 g^*_{\rm d}T_\eta^4/30$, which is conveniently parametrized as an additional number of neutrinos species, $\Delta N_{\nu}\equiv{\rho_{\rm DR}}/{\rho_{\nu}}=(4/7)g^*_{\rm d}\left({T_\eta}/{T_\nu}\right)^4$,
using the known energy density $\rho_\nu$ of a single flavor of an active neutrino at temperature $T_\nu$.

The DM density is set by its chemical freeze-out.  In the regime of our interest (where $\Delta m_\chi \ll m_\chi$) the DM chemical freeze-out is determined by the co-annihilation process $\chip\,\chim \to \rho\,\eta$, with the co-annihilation cross section approximately given by\,\cite{Garcia-Cely:2013nin}, $\sv \simeq {\alpha_{\rm d}^2 \pi/m_\chi^2}$,
where $\alpha_{\rm d} = f_{\rm d}^2/(4\pi)$. 
The contribution from all other channels are $p$-wave suppressed and subleading. The observed DM fraction $\Omega_\text{DM}=0.11\,h^{-2}$ is obtained for $\sv\simeq(2-5)\times 10^{-26}\,\text{cm}^3/{\rm s}$\,\cite{Steigman:2012nb}. For simplicity, we neglect this variation here and take $\sv= 3\times 10^{-26}\,\text{cm}^3/{\rm s}$, as an illustrative value. We will find that the model prefers $m_\chi \lesssim$GeV, so that the observed $\Omega_{\rm DM}$ needs $f_{\rm d}\ll1$, which self-consistently motivates a parametric suppression of $\Delta m_\chi \,(\equiv 2f_{\rm d} v_\phi)$.

\begin{figure}[!t]
\centering
\includegraphics[height=1.7cm]{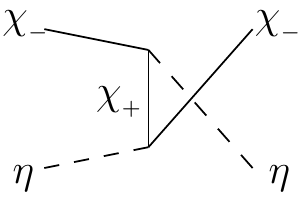}\,
\includegraphics[height=1.7cm]{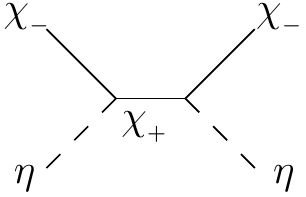}\,
\includegraphics[height=1.7cm]{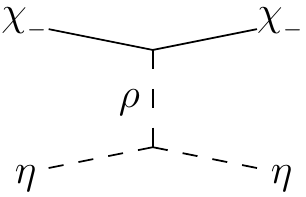}
\caption{DM-DR scattering via $u$, $s$, and $t$ channels.}\label{DMDR:diagram}
\end{figure}

{\bf{\emph{Scattering in the Dark Sector.--}}}  
The DM particle $\chim$ scatters with DR, i.e., the massless pseudoscalar $\eta$, through the processes shown in Fig.\,\ref{DMDR:diagram}.
The cross section for DM-DR scattering is
\begin{equation}
\sigma_{\eta\chim} = {8\pi \alpha_{\rm d}^2 \omega^4 \over \Delta m_\chi^6} \, {\left(1+ { 16\Delta m_\chi^2\over 3m_\rho^2}  + {8\Delta m_\chi^4\over m_\rho^4}\right)}\,, 
\end{equation}
in the limit of $\Delta m_\chi\ll m_\chi$ and where $\omega$ is the energy of $\eta$ in the center-of-mass frame, roughly $T_{\eta}$. One can see that a small $\Delta m_\chi$ enhances the DM-DR scattering.

DM remains in kinetic equilibrium with DR until the momentum exchange rate due to this process becomes smaller than the Hubble expansion rate\,\cite{Boehm:2000gq, Hofmann:2001bi, Green:2003un, Green:2005fa}, i.e., 
$
(T_\eta/m_\chi)n_{\eta}\sigma_{\eta\chim}\sim H(T_\gamma)\,,
$
where $n_\eta = 3\zeta(3)T_{\eta}^3/(4\pi^2)$ is the DR number density. The above condition determines the temperature of kinetic decoupling, $T_{\rm kd}\equiv T_\gamma|_{\rm kd}$. We have
\begin{equation}
T_{\rm kd}\simeq 0.5\text{\,keV} \frac{\delta}{10^{-4.5}} \bigg({m_\chi \over \text{GeV}} \bigg)^{7/ 6} \bigg({10^{-4}\over \alpha_{\rm d}}\bigg)^{1/3} {\xi_{\rm kd}^{-{4/3}}},
\end{equation}
where $\delta\equiv \Delta m_\chi/m_\chi$ is the fractional mass difference, and $\xi_{\rm kd}\equiv(T_\eta/T_\gamma)|_{\rm kd}$, which is in the range of $(0.5-0.8)$. DM-DR scattering can lower $T_{\rm kd}$, which enhances the acoustic damping cutoff, $M_\text{cut}$, in the structure power spectrum. Quantitatively\,\cite{Loeb:2005pm}, we have $M_{\rm cut} \simeq 1.7\times 10^8\left(T_{\rm kd}/\text{keV}\right)^{-3}\,M_\odot$, such that $T_{\rm kd}\simeq$ 0.5\,keV suppresses formation of DM protohalos smaller than about $10^9 M_{\odot}$, and eases the \emph{missing satellites} problem\,\cite{Bringmann:2006mu}. It is important to note that the acoustic damping cut-off in our model is different in shape (oscillatory, with power-law envelope)~\cite{Boehm:2014vja, Buckley:2014hja}, from the free-streaming cut-off in warm dark matter models (exponential). As a result, Lyman-$\alpha$ and other constraints on warm dark matter~\cite{Schneider:2013wwa, Viel:2013fqw} cannot be applied directly~\cite{Boehm:2014vja,Buckley:2014hja}.


DM particles can scatter with each other via $\chim\chim\leftrightarrow\chim\chim$, mediated by the scalar $\rho$. The $t$ and $u$ channel amplitudes dominate the self-scattering due to the smallness of the energy transfer.
In the nonrelativistic limit, the squared-matrix element is approximately
$|i{\emph M}|^2 \simeq {4m_\chi^4/m_\rho^4}$, and the interaction is better described by an attractive Yukawa potential 
$V(r) = -{(\alpha_{\rm d} / r) } e^{ - r\,m_\rho}$,
whose cross section has been studied extensively in the literature \cite{Khrapak:2003, Feng:2009hw, Tulin:2013teo}. For instance, in the Born regime, $\alpha_{\rm d} m_\chi /m_\rho \ll 1$, the scattering cross section in the center of
mass frame, $\sigma_T = \int\! \, d\Omega\,({d\sigma}/{d\Omega}) (1 - \cos\theta) $, is given  by\,\cite{Feng:2009hw}
\begin{align}
  \sigma_T \simeq \frac{8\pi\alpha_{\rm d}^2}{m_\chi^2 v_\text{rel}^4}
                  \bigg[ \log(1 + R^2) - \frac{R^2}{1 + R^2} \bigg] \,,
  \label{eq:sigmaT}
\end{align}
with $R \equiv v_\text{rel} m_\chi  / m_\rho $. Here, $v_\text{rel}$ is the relative
velocity of the two colliding DM particles. We use $\sigma_T$ in the nonperturbative and resonant regimes from Ref.\,\cite{Tulin:2013teo}.

The self-scattering of DM can address the other two small-scale problems of DM.
N-body simulations show that $\langle\sigma_T\rangle / m_\chi\sim (0.1-1)\,\text{cm}^2/\text{g}$ at $v_\text{rel}\sim10{\,\rm km/s}$ leads to smoothening of the inner $\sim1\,$kpc of DM halos in dwarf galaxies and mitigates the \emph{cusp vs. core} problem\,\cite{Loeb:2010gj, Vogelsberger:2012ku}. Here, ${\langle...\rangle}$ denotes an average over the velocity distribution, which we take to be of the Maxwell-Boltzmann form with dispersion $v_\text{rel}$. The same effect also tends to make the inner region of dwarf-sized dark matter halos less dense, and is able to alleviate the \emph{too big too fail} problem\,\cite{Vogelsberger:2012ku, Zavala:2012us}. On the other hand, observations of colliding galaxy clusters, e.g., the Bullet cluster, do not show evidence for DM-DM interactions and thus require $\langle\sigma_T\rangle / m_\chi < 1\,{\rm cm^2/g}$ at typical velocities of $v_\text{rel}\sim10^3{\,\rm km/s}$ therein\,\cite{Fox:2009in}. An ab initio calculation, including the above effects, will provide a more detailed and quantitative prediction of the small-scale structure.

\begin{figure}[!t]
\centering
\includegraphics[height=7.1cm]{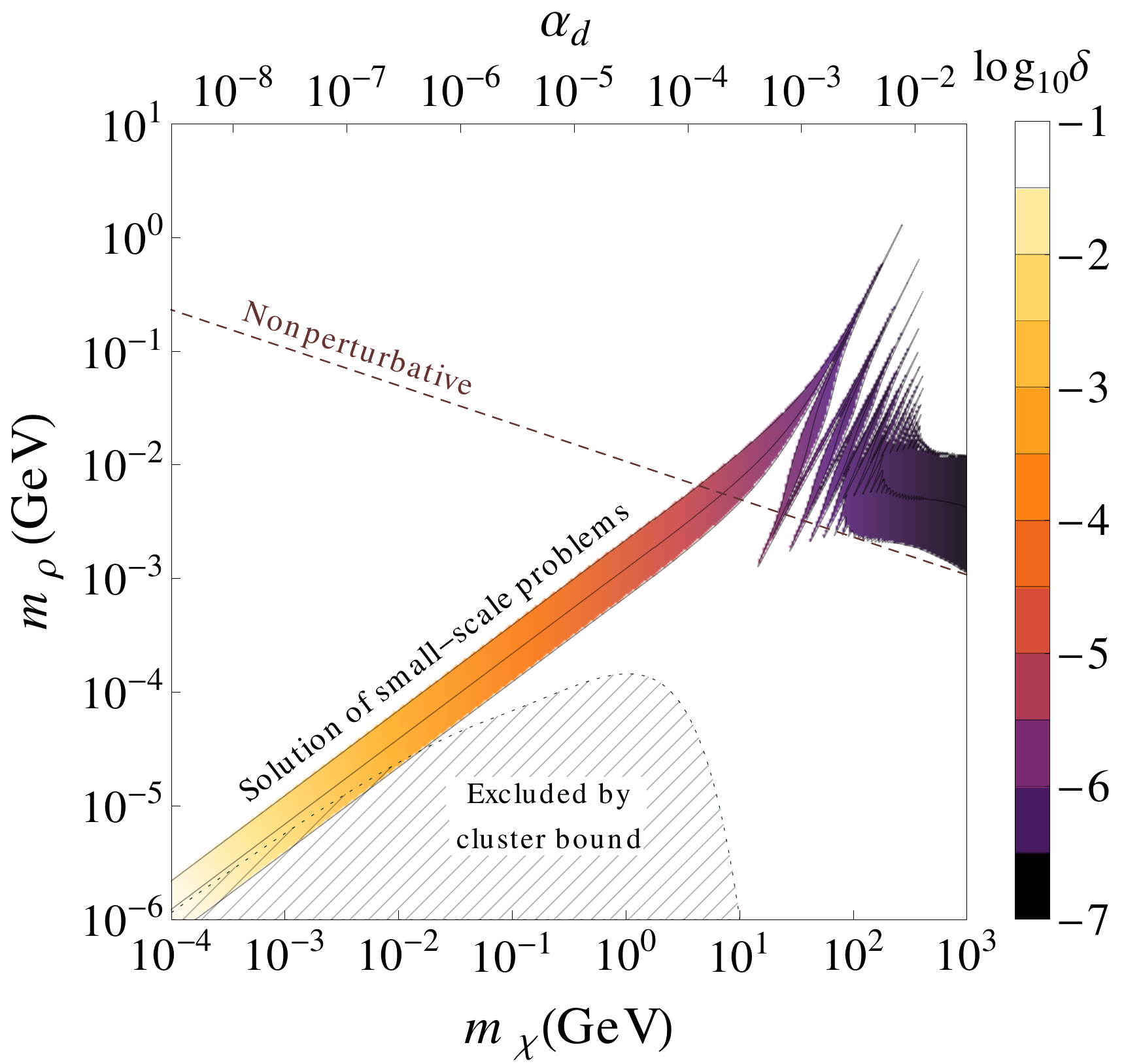}
\caption{Solution of all small-scale problems of DM. For a DM mass $m_\chi$, the coupling $\alpha_{\rm d}$ (top x-axis) is determined by the relic density. Small-scale problems are solved within the band. Three thin solid lines in the band correspond to $\langle\sigma_T\rangle/m_\chi \sim $ 0.1,\,1,\,10\,$\text{cm}^2/\text{g}$, respectively (top-down), at $v_{\rm rel}\sim10\,{\rm km/s}$, addressing the \emph{cusp vs. core} and \emph{too big to fail} problems. The color-gradient inside shows the common logarithm of $\delta\equiv\Delta m_\chi/m_\chi$, which leads to $T_{\rm kd}=0.5\,{\rm keV}$ and solves the \emph{missing satellites} problem. The hatched region at the bottom shows the constraint from galaxy clusters, $\langle\sigma_T\rangle/m_\chi \lesssim 1\,{\rm cm^2/g}$ at $v_{\rm rel}\sim10^3\,{\rm km/s}$, while the dashed line indicates the largest $m_\rho$ for which the scalar potential for $\phi$ is perturbative.}
\label{All:problem}
\end{figure}

In Fig.\,\ref{All:problem}, we show the parameter space where the observed DM abundance is obtained through its chemical freeze-out and all the small-scale problems of DM structure-formation are solved simultaneously. For a given DM mass, the relic density fixes $\alpha_{\rm d}$ (shown on the top x-axis). To solve the \emph{cusp vs. core} and \emph{too big to fail} problems, one needs  $\langle\sigma_T\rangle/m_\chi\sim (0.1-1)\,{\rm cm^2/g}$ at $v_{\rm rel}\sim10\,{\rm km/s}$, which determines $m_\rho$ in terms of $m_\chi$ (shaded band). The oscillatory behavior comes from resonances in the DM scattering. To mitigate the \emph{missing satellites} problem the DM-DR scattering must be enhanced through a near-degeneracy of the masses of $\chi_{\pm}$. The corresponding value of $\delta$, which leads to $T_{\rm kd}={\rm 0.5\,keV}$ and hence to $M_{\rm cut}\simeq10^9\,M_\odot$, is shown through a color-gradient inside the band. We also show the constraint from the galaxy clusters that $\langle\sigma_T\rangle / m_\chi < 1\,{\rm cm^2/g}$ for $v_\text{rel}\sim10^3{\,\rm km/s}$ (hatched region at the bottom is ruled out), and that the theory for the complex scalar $\phi$ is perturbative, i.e., $\lambda_\phi/(4\pi) \lesssim 1$ (below dashed line). 

Some comments are in order about the scenario we identify above. If $m_\chi \lesssim 100\,{\rm keV}$, DM is no longer a truly cold relic, and $\delta$ is no longer small. Also, $m_\rho$ becomes light enough that it directly contributes to DR around $T_{\rm kd}$. On the other hand, if $m_\chi \gtrsim 10\,{\rm GeV}$ the potential for $\phi$ is no longer perturbative. We therefore find $100\,{\rm keV} \lesssim m_\chi\lesssim 10\,{\rm GeV}$ to be best-motivated.

{\bf{\emph{Predictions and Constraints.--}}}
The previous considerations show that the complete scenario is specified in terms of the DM mass if we require that the small-scale problems of DM be resolved. This has two interesting and generic consequences --  (i) DR leads to an observable prediction for $\Delta N_\nu$, and (ii) $m_\chi$ is predicted in the 100\,keV - 10\,GeV range, which may be testable at colliders and direct detection experiments aimed at light DM.

\begin{figure}[!t]
\centering
\includegraphics[width=8.4cm]{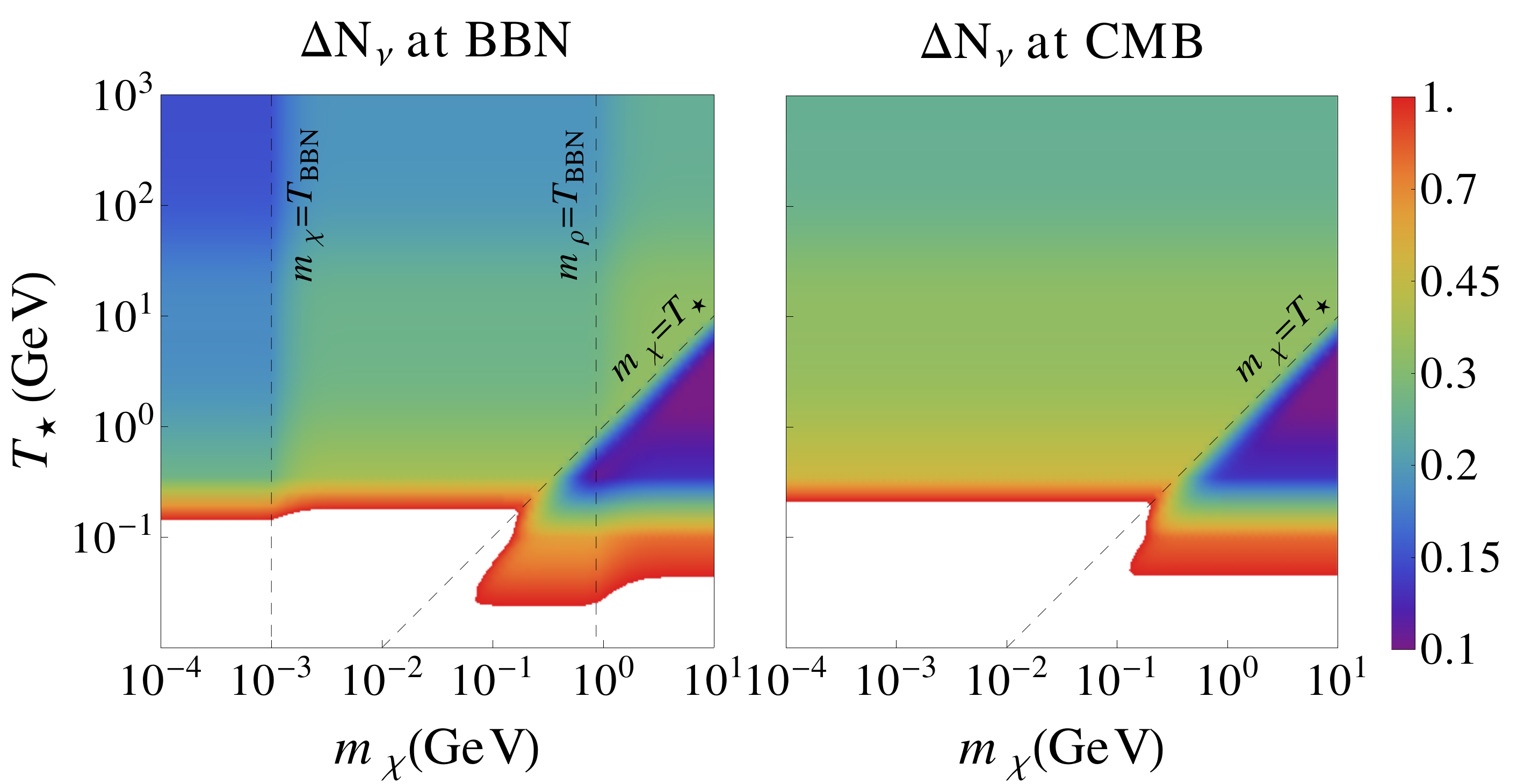}
\vspace{-.4cm}
\caption{$\Delta N_\nu$ generated by dark radiation, depending on the DM mass $m_\chi$ and the temperature $T_\star$ at which the dark and visible sectors decoupled. The white area denotes $\Delta N_\nu \geq 1$. 
}
\label{Neff:kd}
\end{figure}

Fig.\,\ref{Neff:kd} shows the prediction for $\Delta N_\nu$ in BBN and CMB epochs. $\Delta N_\nu$ increases with a later decoupling, i.e., lower $T_\star$, because the decaying SM particles heat up the dark sector as well. This is most apparent for $T_\star\lesssim0.2\,{\rm GeV}$, i.e., below the QCD crossover. There is also a signature \emph{step-like} dependence on $m_\chi$, because of energy injection in the dark sector from $m_\chi$ and $m_\rho$ decays. If $m_\chi > T_\star$, then DM freeze-out heats up both the visible and dark sectors and effectively lowers $\Delta N_\nu$. On the other hand, if $m_\chi$ and/or $m_\rho\simeq10^{-3}m_\chi\gtrsim T_{\rm BBN}\simeq1\,{\rm MeV}$, then the DR bath is heated up by their annihilations/decays, increasing $\Delta N_\nu$ at BBN. These effects lead to sharp changes in $\Delta N_\nu$ across the mass thresholds. Such an effect also exists for $\Delta N_\nu$ at CMB, however for much smaller masses of $m_\chi$ and $m_\rho$ not relevant here. Note also, that there is a minimum DR abundance predicted, i.e., $\Delta N_\nu\simeq0.13$. Dark acoustic oscillations may also constrain this scenario\,\cite{Cyr-Racine:2013fsa}.

The connection between the dark sector and the Standard Model (SM) is through the mixing of the dark scalar $\phi$ with the SM scalar doublet $\Phi$, via $\kappa |\phi|^2 |\Phi|^2$. After symmetry breaking, for $m_\rho /m_h \ll 1$, this can be parametrized as a mass-mixing between the dark scalar $\rho$ and the SM Higgs $h$, via a small mixing angle $\theta\simeq\kappa v_\phi v_{SM}/m_h^2$. 

\begin{figure}[!b]
\centering
\includegraphics[width=0.75\columnwidth]{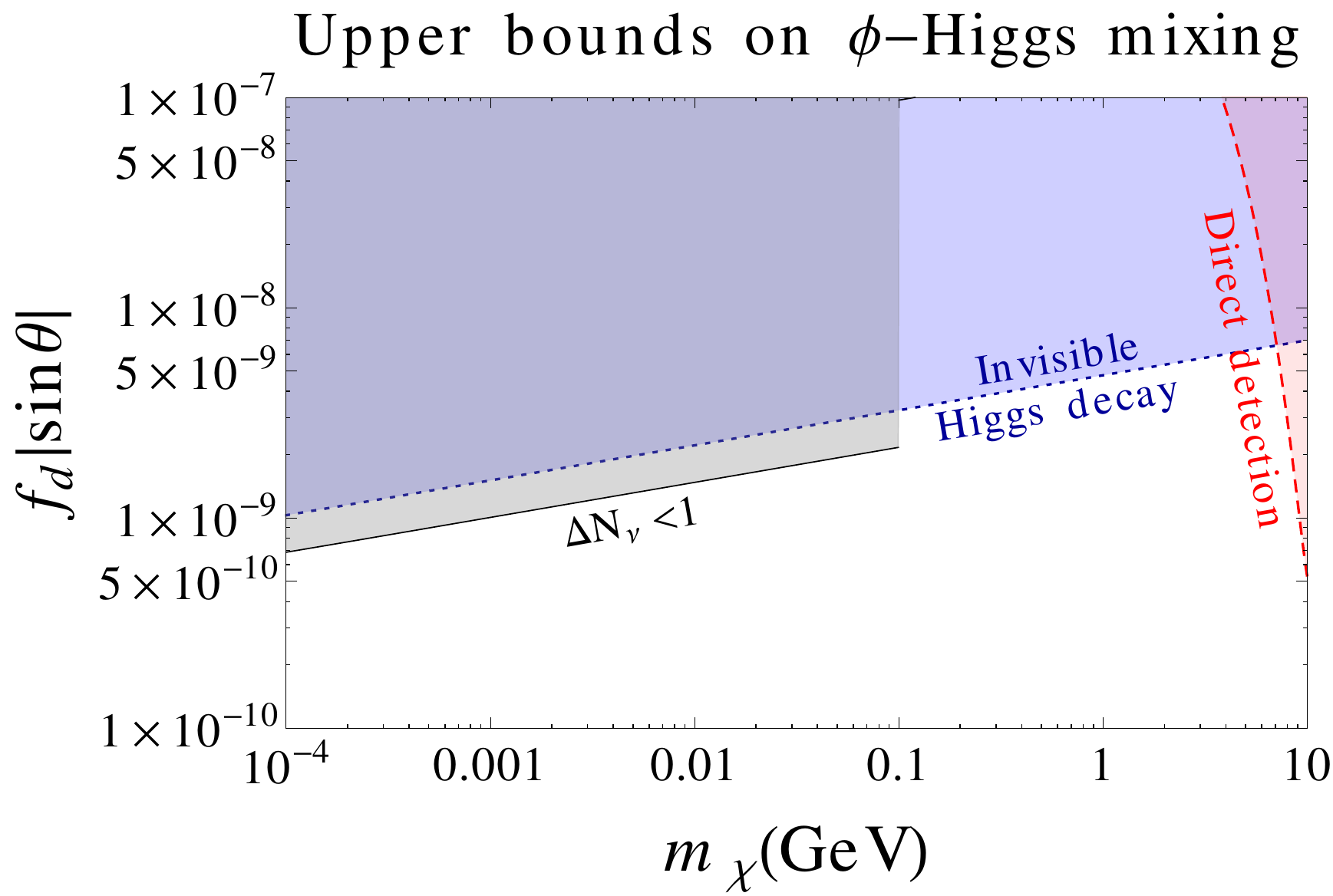}
\vspace{-.4cm}
\caption{Experimental upper bounds on the scalar mixing of $\phi$ and Higgs. Values of $m_\rho$ and $f_{\rm d}$ have been chosen to solve the small-scale problems. The shaded region has been excluded by either DM direct detection (dashed red), invisible Higgs decay (dotted blue), or bounds on $\Delta N_\nu$ (solid gray). }\label{exp:detection}
\end{figure}

The $\phi$-Higgs mixing is probed in three independent and complementary ways, as shown in Fig.\,\ref{exp:detection}. (i) $\Delta N_\nu$ increases due to the thermalization of the dark sector with the visible sector through $\eta\eta\leftrightarrow {\rm SM\,SM}$. This sets an upper bound $\kappa\lesssim 0.01$, if we require $\Delta N_\nu \lesssim 1$\,\cite{Mangano:2011ar, Giusarma:2014zza}.  If the small-scale problems are to be addressed successfully, this translates to $f_{\rm d}|\sin\theta|\,\lesssim \,3.2\times10^{-9}\,(m_\chi/{\rm GeV})^{1/6}$ (gray line). Note that for $m_\chi\gtrsim0.1\,{\rm GeV}$ and $m_\chi>T_\star$, one predicts a smaller $\Delta N_\nu$ as shown in Fig.\,\ref{Neff:kd}, and the bound is weakened significantly. (ii) Invisible decays of the $h$ to $\rho$ or $\eta$ at colliders also probe the same coupling and set an upper bound\,\cite{Belanger:2013kya, Giardino:2013bma} that is translated to $f_{\rm d}|\sin\theta|\lesssim4.7\times10^{-9}\,(m_\chi/{\rm GeV})^{1/6}$\,(blue line). These two bounds are comparable, and in future more precise measurements of Higgs decay at colliders will probe the regime with a sub-GeV $m_\chi$. (iii) The bounds on the DM-nucleon cross section from present experiments, e.g., SuperCDMS and LUX\,\cite{Agnese:2013jaa, Akerib:2013tjd}, are sensitive for $m_\chi$ in the $({\rm few}-10)$\,GeV range, and give an upper limit $f_{\rm d}|\sin\theta|\lesssim 0.5\times10^{-7}$ at $4\,{\rm GeV}$\,(red line), degrading exponentially with smaller DM masses. In the $(7-10)\,{\rm GeV}$ range, direct detection is marginally more sensitive than colliders or cosmology, and may detect a DM signal. Encouragingly, planned direct detection experiments will probe this scenario more sensitively\,\cite{Akerib:2012ys, Yue:2013jja}.

{\bf{\emph{Discussion and Summary .--}}} We have shown that a complex scalar and a fermion charged under a global $U(1)$ symmetry can not only explain the existence and abundance of DM and DR, but also imbue DM with improved scattering properties at galactic scales. This mechanism doesn't require any ad hoc particles or interactions, has the minimal structure required, and is conceptually simpler compared to previous approaches~\cite{Aarssen:2012fx, Dasgupta:2013zpn, Bringmann:2013vra, Ko:2014bka} to this persistent and well-studied problem.

One important aspect of our work is identification of the helicity properties of the scattering amplitudes. It was known that scalar-mediated scattering of DM off a bath of light fermions can't produce late kinetic decoupling\,\cite{Aarssen:2012fx}. A bath of dark massless gauge bosons is also prevented from delaying DM kinetic decoupling to $T\lesssim{\rm keV}$ if ellipticity bounds on galaxies apply\,\cite{Feng:2009mn}. We showed that these limitations can be overcome if the bath is composed of a pseudoscalar instead. The Goldstone mode of the spontaneously broken $U(1)$ provides such a pseudoscalar that delays DM kinetic decoupling, while the scalar mode enhances DM self-scattering. 

Phenomenologically the idea is viable and testable. The specific scenario we identified, i.e., the pseudo-Dirac DM regime of Ref.\,\cite{Weinberg:2013kea}, addresses the small-scale problems of DM, and we explored reheating effects in the dark sector. The main signatures are a light DM, in the $100{\,\rm keV}-10{\,\rm GeV}$ range, and peculiar patterns for $\Delta N_\nu$, e.g., BBN abundances may show a $\Delta N_\nu$ that's smaller than in CMB measurements. This is due to annihilations and decays of $\chim$ and/or $\rho$, as shown in Fig.\,\ref{Neff:kd}. The connection to the SM may be probed at colliders, future direct detection experiments, and using precise cosmological measurements of $\Delta N_{\nu}$, etc.\,\cite{Cyr-Racine:2013fsa}. On the other hand, absence of an appreciable abundance of DR, e.g., measuring $\Delta N_{\nu}\lesssim 0.13$ in the CMB epoch, would rule-out this scenario. The possibility, that the nature of DM and DR and their shared symmetry could be inferred via their consequences on astrophysical structures, and revealed cosmologically, is nonetheless remarkable.

{\bf{\emph{Acknowledgements.--}}} We thank our colleagues at ICTP, John Beacom, Gary Steigman, Michel\,H.\,G.\,Tytgat, and Francis-Yan Cyr-Racine for useful comments. We also acknowledge the use of FeynCalc\,\cite{Mertig:1990an} and JaxoDraw\,\cite{Binosi:2008ig}.

\bibliographystyle{apsrev}
\interlinepenalty=10000
\tolerance=100

\bibliography{./DR-DM}

\end{document}